\documentclass[aps,pra,twocolumn,showpacs,superscriptaddress,nofootinbib]{revtex4-2}
\usepackage{amsmath,amssymb,amsfonts}
\allowdisplaybreaks[4]
\usepackage{graphicx}
\usepackage{dcolumn}
\usepackage{epstopdf}
\usepackage[T1]{fontenc}
\usepackage{newtxtext}
\usepackage{subfigure}
\usepackage[percent]{overpic}
\usepackage{float}
\usepackage{bm}
\usepackage{algorithm}
\usepackage{algpseudocode}
\usepackage{booktabs}
\usepackage{float}
\usepackage{booktabs}

\allowdisplaybreaks

\usepackage[unicode]{hyperref}
\hypersetup{
	unicode=true,          
	a4paper=true,
	plainpages=false,
	pdftitle={Title of PDF},    
	pdfauthor={Author of PDF},     
	pdfsubject={Subject of PDF},   
	colorlinks=true,       
	linkcolor=blue,          
	citecolor=blue,        
	filecolor=blue,      
	urlcolor=blue           
}
\newcommand{\refsec}[1]{Sec.~\ref{#1}}

\newcommand{\refeq}[1]{Eq.~(\ref{#1})}

\newcommand{\reffig}[1]{Fig.~\ref{#1}}

\newcommand{\reffigsub}[2]{Fig.~\hyperref[#1]{\ref{#1}(#2)}}
\newcommand{\reftab}[1]{TABLE ~\ref{#1}}

\DeclareMathAlphabet{\mathcal}{OMS}{cmsy}{m,b}{n,it}

\begin{document}

	\title{An efficient preconditioned conjugate-gradient solver for a two-component dipolar Bose–Einstein condensate}
	
	\author{Weijing Bao}
	\email{baoweijing@nuaa.edu.cn}
	
	\author{Zhenhao Wang}
	
	\author{Jia-Rui Luo}
	
	\author{Kui-Tian Xi}
	\email[Corresponding author: ]{xiphys@nuaa.edu.cn}
	\affiliation{College of Physics, Nanjing University of Aeronautics and Astronautics, Nanjing 211106, China}
	\affiliation{Key Laboratory of Aerospace Information Materials and Physics (NUAA), MIIT, Nanjing 211106, China}
	
	\date{\today}
	
	\begin{abstract}
		We develop a preconditioned nonlinear conjugate-gradient solver for ground states of binary dipolar Bose–Einstein condensates within the extended Gross–Pitaevskii equation including Lee–Huang–Yang corrections. The optimization is carried out on the product-of-spheres normalization manifold and combines a manifold-preserving analytic line search, derived from a second-order energy expansion and validated along the exact normalized path, with complementary Fourier-space kinetic and real-space diagonal (Hessian-inspired) preconditioners. The method enforces monotonic energy descent and exhibits robust convergence across droplet, stripe, and supersolid regimes while retaining spectrally accurate discretizations and FFT-based evaluation of the dipolar term. In head-to-head benchmarks against imaginary-time evolution on matched grids and tolerances, the solver reduces iteration counts by one to two orders of magnitude and overall time-to-solution, and it typically attains slightly lower energies, indicating improved resilience to metastability. We reproduce representative textures and droplet-stability windows reported for dipolar mixtures. These results establish a reliable and efficient tool for large-scale parameter scans and phase-boundary mapping, and for quantitatively linking numerically obtained metastable branches to experimentally accessible states.
	\end{abstract}
	
	\maketitle

	\section{Introduction}
	
	Dipolar Bose-Einstein condensates (BECs) provide a pristine platform for exploring self-organized quantum phases emerging from the competition between short-range contact interactions and long-range, anisotropic dipole-dipole forces. The observation of self-bound quantum droplets in dysprosium \cite{Kadau2016observing} spurred extensive theoretical and experimental activity, with explanations in terms of effective three-body interactions \cite{Xi2016droplet,Bisset2015crystallization} and stabilization by quantum fluctuations in the form of Lee-Huang-Yang (LHY) corrections \cite{Wachtler2016quantum,Saito2016path}. Building on these developments, coordinated studies \cite{Blakie2018droplet,Roccuzzo2019supersolid,Tanzi2019observation} culminated in the first realizations of dipolar supersolidity, reported nearly simultaneously in erbium \cite{Natale2019excitation} and dysprosium \cite{Bottcher2019transient,Tanzi2019supersolid}. In parallel, the creation of a two-component dipolar condensate (Er-Dy) \cite{Trautmann2018dipolar} opened the door to mixture physics, and subsequent work examined supersolidity as well as ground-state and dynamical properties in binary settings \cite{Xi2018figering,Lavoine2021beyond,Li2022long,Bisset2021quantum,Smith2021quantum,Kirkby2024excitation,Bland2022alternating,Zhang2024metastable,Santos2023self,Norcia2021two,Durastante2020feshbach,Politi2022interspecies}. Together these milestones establish dipolar mixtures as a versatile platform for emergent quantum order.
	
	Ground-state computation is central to mapping phase diagrams, assessing stability, and providing faithful initial data for dynamics and quantitative comparison with experiment. Formally, one seeks minimizers of the extended Gross-Pitaevskii energy (eGPE) under normalization constraints. In two-component dipolar gases this task is particularly demanding. Intra- and inter-species couplings enlarge parameter space and enable miscibility transitions and partial phase separation; nonlocal dipolar interactions and beyond-mean-field contributions render the landscape strongly nonconvex with many low-lying minima; and numerically one must handle two coupled fields, multiple constraints, and FFT-based dipolar convolutions with high-order (often spectral) discretizations. As a consequence, simple gradient-flow schemes may converge slowly and are prone to becoming trapped in metastable states, especially in droplet and supersolid regimes.
	
	Imaginary-time evolution (ITE) and related gradient-flow approaches are widely used for the eGPE \cite{Antoine2014gpelab,Bao2015ground,Bao2004ground,Lee2016phase,Pyzh2020phase,Kumar2015fortran,Xi2018figering,Lavoine2021beyond,Li2022long,Bisset2021quantum,Smith2021quantum,Kirkby2024excitation,Bland2022alternating,Zhang2024metastable,Santos2023self}, typically in combination with backward-Euler finite differences (BEFD) \cite{Bao2004computing,Bao2004ground}, sine-pseudospectral variants (BESP/BFSP) \cite{Bao2006efficient,Bao2010efficient}, time-splitting spectral schemes (TSSP/SSFM) \cite{Bao2004computing}, or, less commonly, Crank-Nicolson methods \cite{Qing2006computing,Vudragovic2012c}. Beyond gradient flow which is viewed as projection-based constrained minimization, Riemannian optimization on the normalization manifold offers an alternative class of methods. Steepest-descent and conjugate-gradient (CG) iterations that preserve normalization by construction are well developed in numerical optimization and for PDE constrained problems \cite{Hager2006survey,Absil2009optimization}. For single-component condensates, Antoine and co-workers showed that preconditioned CG can markedly outperform ITE in efficiency and robustness \cite{Antoine2017efficient}, and subsequent studies incorporated dipole–dipole interactions and LHY corrections \cite{Antoine2018preconditioned,Hertkorn2024self,Hertkorn2021pattern,Shu2024preconditioned,Zhang2024new}. A dedicated, systematically preconditioned CG framework for a two-component dipolar condensate, however, remains to be developed.
	
	In this paper we close that gap by developing a \emph{preconditioned nonlinear conjugate-gradient} (CG) solver for binary dipolar Bose-Einstein condensates including Lee-Huang-Yang (LHY) corrections. The minimization is formulated on the \emph{product-of-spheres normalization manifold}, so the two component norms are enforced by construction via projected directions. Building on this geometry, we derive a manifold-preserving analytic line search from a quadratic energy model that explicitly includes cross-component mean-field terms, the nonlocal dipolar interaction, and LHY variations, and we validate the proposed step along the exact normalized path. To address the coupled stiffness specific to mixtures, we introduce \emph{complementary preconditioners}, namely a \emph{real-space diagonal (Hessian-inspired)} scaling that captures trap/contact/LHY curvature and a \emph{Fourier-space kinetic} scaling that damps high-$k$ modes, applied in sequence to balance low- and high-frequency stiffness. We further employ a \emph{robust direction update} (Hestenes–Stiefel $\beta$ with restarts and a monotonicity safeguard that reverts to preconditioned steepest descent when the quadratic model is unreliable), which is essential in multistable droplet/stripe/supersolid regimes.
	
	Compared with CG schemes for single-component condensates, the present method introduces a mixture-aware geometric formulation that updates in the tangent space of the product-of-spheres manifold, thereby enforcing two normalization constraints and consistently accounting for interspecies couplings. It further provides an analytic, manifold-preserving line search for mixtures that incorporates cross-component mean-field terms, the nonlocal dipolar interaction, and LHY corrections in closed form. Preconditioning is tailored to mixtures by composing a real-space diagonal (Hessian-inspired) scaling with a Fourier-space kinetic scaling to control low- and high-$k$ stiffness across scales. Finally, robustness is ensured by algorithmic safeguards, including Hestenes-Stiefel $\beta$ with periodic restarts and a $\beta \to 0$ fallback, which maintain monotone energy descent in highly nonconvex, metastable landscapes.
	
	In head-to-head benchmarks against imaginary-time evolution (ITE) across representative regimes, including droplet, stripe, and supersolid phases, the proposed solver achieves robust, monotonic energy decay, reduces iteration counts by one to two orders of magnitude at matched tolerances, and typically attains lower energies, translating into a clear time-to-solution advantage despite a costlier per-step transform budget. We also reproduce key mixture results in \cite{Santos2023self,Zhang2024metastable}, underscoring the physical reliability of the method. Implementation details (discretization, dipolar convolution, FFT accounting, and stopping criteria) are documented to ensure reproducibility and to facilitate large-scale parameter scans and phase-boundary mapping in multi-component dipolar gases.
	
	The paper is organized as follows. In \refsec{section:formalism} we define the energy functional and coupled Gross-Pitaevskii equations for a two-component dipolar condensate. The CG procedure, including pseudocode and implementation details, is presented in \refsec{section:algorithm}. A hybrid update strategy for the conjugate direction is described in \refsec{section:Conjugate Direction}. In \refsec{section:Optimal step length} we introduce the analytic line search derived from a quadratic energy expansion. Numerical results, including comparisons among preconditioners and between CG and ITE, reproductions of key results in \cite{Santos2023self,Zhang2024metastable}, are reported in \refsec{section:Numerical Results}. We conclude in \refsec{section:Conclusion}.
	
	\section{Formalism}
	\label{section:formalism}
	
	We consider a two-component dipolar Bose-Einstein condensate (BEC) with Lee-Huang-Yang (LHY) correction. Throughout, we work in oscillator units with characteristic frequency $ \omega_{0} $: length $ l = \sqrt{\hbar / (m \omega_{0})} $ and energy $ \hbar \omega_{0} $. The component wave functions are normalized as $ \int |\psi_{i}(\bm{r})|^{2} d\bm{r} =1 $, and $ N_{i} $ enter the couplings explicitely. In these units, the dimensionless energy functional reads
	\begin{align}
		E[\psi_{1,2}]
		= \int d\bm{r} \Bigg\{  \sum_{i=1}^{2} N_i \bigg[ & \frac{1}{2} |\nabla\psi_i|^2  + V_{\text{ext}}(\bm{r}) |\psi_i|^2 \nonumber\\
		& + \frac{1}{2} \sum_{j=1}^{2}   G_{ij} |\psi_{i}|^2 |\psi_{j}|^2 \nonumber\\
		& + \frac{1}{2} \sum_{j=1}^{2}   D_{ij}|\psi_{i}|^2 \nonumber\\
		& \times \int d\bm{r^{\prime}} \frac{1 - 3 \cos^{2}\theta}{\left| \bm{r} - \bm{r^{\prime}} \right|^3} |\psi_j|^2 \bigg] \nonumber \\
		& + \frac{\sqrt{2}}{15\pi^2} \int_0^1 du \sum_{\lambda=\pm} {V_\lambda(u)}^{5/2} \Bigg\}.
		\label{eq:dimensionless form of the energy functional}
	\end{align}
	The dimensionless harmonic trap potential is 
	\begin{equation}
		V_{\text{ext}}(\bm{r}) = \frac{1}{2} \lambda_{x}^{2} x^2 + \frac{1}{2} \lambda_{y}^{2} y^2 + \frac{1}{2} \lambda_{z}^{2} z^2,
	\end{equation}
	with $ \lambda_{x} = \omega_{x} / \omega_{0} $, $ \lambda_{y} = \omega_{y} / \omega_{0} $ and $ \lambda_{z} = \omega_{z} / \omega_{0} $. 
	$G_{ij} = 2\pi a_{ij} N_j / l$ is the contact interaction coefficient with the s-wave scattering length $ a_{ij} $, $ (i, j = 1, 2) $. The dipole-dipole interaction coefficient is given by
	$D_{ij} = \mu_{0} \mu_{i} \mu_{j} N_{j} / (4\pi l^{3} \hbar \omega_{0})$ with the vacuum magnetic permeability $ \mu_0 $ , and the magnetic dipole moment $\mu_i$.
	In the LHY correction term,
	\begin{align}
		V_\pm (u) &= \sum_{j=1}^2 \eta_{jj} n_j \pm \sqrt{\delta^2 + 4\eta_{12} \eta_{21} n_1 n_2},
	\end{align}
	with $ \delta = \eta_{11}n_1 - \eta_{22} n_2 $, and $ \eta_{ij} = G_{ij} + D_{ij} (3u^2 - 1) $, where $ u=\cos \theta $ with $ \theta $ the angle between the relative position vector of the two dipoles and their polarization direction. We denote the component densities by $n_i(\bm r) \equiv |\psi_i(\bm r)|^2$.
	
	The dimentionless Gross-Pitaevskii equations is given by
	\begin{align}
		i \frac{\partial \psi_{i}}{\partial t} 
		= &\bigg[ -\frac{1}{2} {\nabla}^{2} + V_{\text{ext}}(\bm{r})  + \sum_{j=1}^{2} G_{ij} \left| \psi_{j} \right|^{2}  \nonumber \\ 
		& + D_{ij}\int d\bm{r^{\prime}} \frac{1 - 3 \cos^{2}\theta}{\left| \bm{r} - \bm{r^{\prime}} \right|^3} |\psi_j|^2 \nonumber \\ 
		& + \mu_{\text{LHY}}^{(i)} \left[n_{1,2}(\bm{r})\right]\bigg] \psi_{i} ,
	\end{align}
	with the LHY correction
	\begin{align}
		\mu_{\text{LHY}}^{(i)}\left[n_{1,2}(\bm{r})\right] = &\frac{1}{3 \sqrt{2} \pi^2 N_i} \int_0^1 du \sum_{\lambda=\pm} {V_\lambda (u)}^{3/2}  \nonumber 
		\\ &\times \left[\eta_{ii} \pm \frac{(-1)^{3-i} \eta_{ii} \delta + 2\eta_{12} \eta_{21} n_{3-i}}{\sqrt{\delta^2 + 4\eta_{12} \eta_{21} n_1 n_2}} \right]. 
		\\ &\quad \nonumber
	\end{align}
	The Hamiltonian of each component is 
	\begin{align}
		\bm{\hat{H}}_i^{(n)}(\bm{r}) = & -\frac{1}{2} {\nabla}^{2} +  V_{\text{ext}} (\bm{r}) + \sum_{j=1}^{2} G_{ij} \left|\psi_{j}(\bm{r}) \right|^{2}  \nonumber \\ 
		&+ \sum_{j=1}^{2} D_{ij} \int d\bm{r^{\prime}} \frac{1 - 3 \cos^{2}\theta}{\left| \bm{r} - \bm{r^{\prime}} \right|^3} |\psi_{j}(\bm{r^{\prime}})|^2 \nonumber \\ 
		&+ \mu_{\text{LHY}}^{(i)} \left[n_{1,2}(\bm{r})\right].
	\end{align}

	\section{Algorithm}
	\label{section:algorithm}
	
	The conjugate-gradient (CG) approach has been successfully employed for single-component dipolar BECs with LHY corrections \cite{Hertkorn2024self,Hertkorn2021pattern,Antoine2017efficient,Antoine2018preconditioned,Shu2024preconditioned,Zhang2024new}. Here we extend it to binary dipolar condensates within the eGPE framework. We treat the minimization as a constrained optimization on the product-of-spheres manifold $ \mathcal{M} = { \psi_{1}, \psi_{2}: \|\psi_{i}\|_{2} = 1 } $. At iteration $ n $, the Lagrange multipliers (chemical potentials) are chosen as 
	\begin{equation}
		\mu_{i}^{(n)} = \langle \psi_{i}^{(n)}, \hat{\bm{H}}_{i} [\psi_{1}^{(n)}, \psi_{2}^{(n)}] \psi_{i}^{(n)} \rangle,
	\end{equation}
	so that the Euclidean gradient $ g_{i}^{(n)} = \hat{\bm{H}}_{i} \psi_{i}^{(n)} - \mu_{i}^{(n)}\psi_{i}^{(n)}$ is automatically orthogonal to $ \psi_{i}^{(n)} $ up to round-off. The Riemannian (tangent) residual is taken as the projection 
	\begin{equation}
		r_{i}^{(n)} = \text{Proj}_{\psi_{i}^{(n)}} [g_{i}^{(n)}] = g_{i}^{(n)} - \text{Re} \langle g_{i}^{(n)}, \psi_{i}^{(n)} \rangle \psi_{i}^{(n)} = g_{i}^{(n)},
	\end{equation}
	and we precondition it with a block-diagonal operator $ \mathcal{P}^{-1} = \text{diag} (\mathcal{P}_{1}^{-1}, \mathcal{P}_{2}^{-1}) $. We use two complementary choices: a kinetic preconditioner $ \mathcal{P}_{\Delta, i}^{-1} $ that is diagonal in Fourier space, and a diagonal (Hessian-inspired) preconditioner $ \mathcal{P}_{V,i}^{-1} $ that is diagonal in real space; weighted combinations are possible. The search direction is updated by a Hestenes–Stiefel (HS)–type parameter $ \beta^{(n)} $ with periodic restarts to maintain descent. A manifold-preserving line search updates both components along the great-circle geodesic 
	\begin{equation}
		\psi_{i}^{(n)} (\theta) = \cos \theta \psi_{i}^{(n)} + \sin \theta \hat{p}_{i}^{(n)},
	\end{equation}
	where $ \hat{p}_{i}^{(n)} = p_{i}^{n} / \| p_{i}^{(n)} \|_{2} $, and $ \theta $ is chosen by an analytic quadratic model of the total energy $ E[\psi_{1}, \psi_{2}] $. As a safeguard when the quadratic model is unreliable (e.g., negative curvature or near cancellation), we set $ \beta^{(n)} = 0 $ and restart the direction, reducing the update to a (preconditioned) steepest-descent step.
	
	\begin{algorithm}[H]
		\caption{The conjugate gradient method.}\label{alg:cg}
		\begin{algorithmic}
			\While{not converged}
			\State $r_i^{(n)} = \bm{\hat{H}}_i^{(n)} \psi_i^{(n)} - \mu_i \psi_i^{(n)}$
			\State $d_i^{(n+1)} = - (\mathcal{\bm{P}}^{-1})^{(n+1)} r_i^{(n+1)}+ \beta^{(n+1)} p_i^{(n)}$
			\State $p_i^{(n)} = d_i^{(n)} - \text{Re} \left\langle d_i^{(n)} \bigg| \psi_i^{(n)} \right\rangle \psi_i^{(n)}$
			\State $\theta_i^{(n)} = \arg\min_{\theta} E \left( \cos(\theta) \psi_i^{(n)} + \sin(\theta) p_i^{(n)} / \| p_i^{(n)} \| \right)$
			\State $\psi_i^{(n+1)} = \cos(\theta_i^{(n)}) \psi_i^{(n)} + \sin(\theta_i^{(n)}) p_i^{(n)} / \| p_i^{(n)} \|$
			\State $n = n + 1$
			\EndWhile
		\end{algorithmic}
	\end{algorithm}
	
	\section{Conjugate Direction}
	\label{section:Conjugate Direction}
	
	\subsection{Coupled Residual}
	
	At iteration $ n $, the stationary states satisfy $\bm{\hat{H}}_i^{(n)} \psi_i^{(n)} = \mu_i^{(n)} \psi_i^{(n)} $, where $\bm{\hat{H}}_i^{(n)}$ is the mean-field Hamiltonian (including dipolar and LHY contributions at the current iterate) and $\mu_i^{(n)} = \langle \psi_i^{(n)} | \bm{\hat{H}}_i^{(n)} | \psi_i^{(n)} \rangle$ are the component chemical potentials.
	
	The coupled Hamiltonians, chemical potentials, and wave functions can be written as follows, respectively,
	\begin{align}
		\bm{\hat{H}}^{(n)} &= 
		\begin{bmatrix} 
			\bm{\hat{H}}_1^{(n)}&\quad&0 \\ 0 &\quad& \bm{\hat{H}}_2^{(n)}
		\end{bmatrix}, \\ 
		\mu^{(n)} &= 
		\begin{bmatrix} 
			\mu_1^{(n)}&\quad&0 \\ 0 &\quad& \mu_2^{(n)} 
		\end{bmatrix}, \\ 
		\psi^{(n)} &= \frac{1}{N} 
		\begin{bmatrix} 
			N_1  \psi_1^{(n)}\\ N_2 \psi_2^{(n)} 
		\end{bmatrix},
	\end{align}
	with $N = N_1+N_2$. Throughout, we enforce the normalization constraints $ \| \psi_{i}^{(n)} \|_{2} = 1 $ (product-of-spheres manifold).
	The coupled residual is given by
	\begin{equation}
		r^{(n)}
		= \bm{\hat{H}}^{(n)} \psi^{(n)} - \mu^{(n)} \psi^{(n)}.
	\end{equation}
	It ensures $ \text{Re} \langle r_{i}^{(n)}, \psi_{i}^{(n)}  \rangle = 0 $, i.e., the search stays on the normalization manifold.

	\subsection{Preconditioning Matrix}
	\label{section:Preconditioning MAtrix}
	
	Following the approach of Ref. \cite{Antoine2017efficient}, we employ preconditioning to accelerate convergence of the constrained minimization while preserving robustness near multistable regimes. In our setting, the preconditioner acts on the coupled residual in a block-diagonal fashion (one block per component), so that application remains inexpensive and does not require solving fully coupled linear systems at each iteration. The guiding principle is to approximate the dominant local curvature of the energy at the current iterate while keeping the cost limited to pointwise operations or, at most, one FFT pair per component.
	
	A natural choice is the \emph{potential energy preconditioner}, constructed from the potential terms of the Hamiltonian. To avoid excessively large matrix elements and maintain numerical stability, we retain the dominant diagonal contributions and incorporate higher-order off-diagonal effects through suitable local averages. This yields the block structure
	\begin{align}
		(\mathcal{\bm{P}}_V^{-1})^{(n)} = 
		\begin{bmatrix} 
			(\mathcal{\bm{P}}_{V1}^{-1})^{(n)} & 0 \\ 
			0 & (\mathcal{\bm{P}}_{V2}^{-1})^{(n)} 
		\end{bmatrix},
	\end{align}
	where
	\begin{align}
		\mathcal{\bm{P}}_{Vi}^{(n)} =  \int d\bm{r} &\bigg( \frac{1}{2} \left| \nabla \psi_j^{(n)} \right|^{2} + V_{\text{ext}} (\bm{r})  \left|\psi_j^{(n)} \right|^{2} + \sum_{j=1}^{2} G_{ij} \left|\psi_j^{(n)} \right|^{4} \nonumber \\
		&+ \mu_{\text{LHY}}^{(i)} \left[n_{1,2}(\bm{r})\right]  \left|\psi_j^{(n)} \right|^{2} \bigg) + V_{\text{ext}}(\bm{r})  \nonumber \\
		&+ \sum_{j=1}^{2} G_{ij}|\psi_{j}^{(n)}|^2 + \mu_{\text{LHY}}^{(i)} \left[n_{1,2}(\bm{r})\right].
	\end{align}
	Operationally, $ (\bm{P}_{V_{i}}^{-1})^{(n)} $ acts as a real-space diagonal operator that rescales the residual by a positive local proxy of curvature assembled from the trap, contact interactions, and the LHY Jacobian evaluated at the current iterate. In practice, we include a small positive shift to regularize near-vanishing local curvature and prevent blowup of the inverse preconditioner. With the sign-indefinite dipolar contribution excluded, the remaining terms, namely the trap potential, contact interactions, and LHY Jacobian, render the operator strictly positive definite; the shift merely strengthens this property and does not alter the fixed points. Because the action is pointwise in real space, the application cost is $O(N)$ per component and is negligible compared with the FFT-based parts of the gradient.
	
	As a complementary strategy, we use the \emph{kinetic-energy preconditioner}, which isolates the dispersive stiffness associated with the Laplacian. By retaining the kinetic term in momentum space, this choice efficiently damps high-$ k $ components and is particularly effective when kinetic contributions dominate (e.g., for fine grids or near roton softening). The block form is
	\begin{align}
		(\mathcal{\bm{P}}_{\Delta}^{-1})^{(n)} = 
		\begin{bmatrix} 
			(\mathcal{\bm{P}}_{\Delta1}^{-1})^{(n)} & 0 \\ 
			0 & (\mathcal{\bm{P}}_{\Delta2}^{-1})^{(n)} 
		\end{bmatrix},
	\end{align}
	where
	\begin{align}
		\mathcal{\bm{P}}_{\Delta i}^{(n)} =  \int d\bm{r}  &\bigg( \frac{1}{2} \left| \nabla \psi_j^{(n)} \right|^{2} + V_{\text{ext}} (\bm{r})  \left|\psi_j^{(n)} \right|^{2} + \sum_{j=1}^{2} G_{ij} \left|\psi_j^{(n)} \right|^{4} \nonumber \\
		&+ \mu_{\text{LHY}}^{(i)} \left[n_{1,2}(\bm{r})\right]  \left|\psi_j^{(n)} \right|^{2} \bigg) + \frac{1}{2} \nabla^2 .
	\end{align}
	In practice, this corresponds to one FFT/iFFT pair per component with a diagonal multiplier in $k$-space, so the asymptotic cost is $ O(NlogN) $ and comparable to a single nonlocal convolution already required by the dipolar term.
	
	It is noteworthy that the two preconditioners can be used together within the same iteration. In accordance with our implementation, we apply them in sequence through the multiplicative composition
	\begin{equation}
		(\mathcal{\bm{P}}^{-1})^{(n)} = (\mathcal{\bm{P}}_{\Delta}^{-1})^{(n)} (\mathcal{\bm{P}}_{V}^{-1})^{(n)}.
	\end{equation}
	This ordering first regularizes the high-frequency content via the kinetic factor and then rescales the residual by the local (potential/Hessian-like) curvature, yielding a balanced reduction of stiffness across scales. Both factors are symmetric positive-definite under mild shifts, so their composition remains well-conditioned at the level relevant for CG updates.
	
	\subsection{Direction Searching Strategy}
	
	The search direction is built from a preconditioned residual and the previous (projected) direction. At the first step a pure preconditioned steepest-descent is taken, while subsequent steps add a conjugate component:
	\begin{align}
		\begin{cases}
			d_i^{(1)} &= - (\mathcal{\bm{P}}_{i}^{-1})^{(1)} r_i^{(1)}, \\
			d_i^{(n+1)} &= - (\mathcal{\bm{P}}_{i}^{-1})^{(n+1)} r_i^{(n+1)} + \beta^{(n+1)} p_i^{(n)},
		\end{cases}
	\end{align}
	where $r_i^{(n)}$ is the (tangent) residual at iteration $n$ and $(\mathcal{\bm{P}}_{i}^{-1})^{(n)}$ is the block preconditioner acting on component $i$. The vector $p_i^{(n)}$ denotes the projected search direction (see \refsec{section:algorithm}) and guarantees tangency to the normalization manifold; in practice this is enforced by subtracting the component parallel to $\psi_{i}^{(n)}$. The inner products below are taken as real parts of the $L^{2}$ products so that global phases do not interfere with conjugacy.
	
	Conjugacy is enforced with respect to the (iteration-dependent) mean-field operator, following a classical nonlinear CG strategy:
	\begin{equation}
		\Big\langle d_i^{(n+1)} \Big| \bm{\hat{H}}^{(n)} \Big| p_i^{(n)} \Big\rangle \equiv 0 ,
		\label{eq:conjugate condition}
	\end{equation}
	where $\bm{\hat{H}}^{(n)}$ serves as a surrogate Hessian at the current iterate and incorporates the couplings between components through the densities that enter the mean-field potentials. Condition \refeq{eq:conjugate condition} ensures that, in the quadratic approximation of the energy, the new direction removes curvature information already captured by $p_i^{(n)}$, thereby improving efficiency over steepest descent.
	
	The scalar $\beta^{(n)}$ controls how much of the previous direction is retained and is critical for robustness in nonconvex landscapes with many low-lying minima. Several choices exist in the literature. A widely used option is the Polak–Ribière$^+$ (PR$^+$) update, which truncates negative values to preserve descent $\beta = \text{max} (\beta_{\text{PR}}, 0)$ as in \cite{Antoine2017efficient,Hertkorn2024self}, with 
	\begin{equation}
		\beta_{\text{PR}}^{(n)} =  \frac{ \text{Re} \langle r^{(n)} - r^{(n-1)}| (\mathcal{\bm{P}}^{-1})^{(n)} | r^{(n)} \rangle }{\text{Re} \langle r^{(n-1)}| (\mathcal{\bm{P}}^{-1})^{(n)} | r^{(n-1)} \rangle}, \quad (n \ge 2).
	\end{equation}
	This formula exploits preconditioned residual correlations across iterations and is inexpensive to evaluate since it reuses quantities already computed for stopping tests.
	
	In strongly multistable regimes, typical of two-component dipolar condensates, residuals may decorrelate, and PR$^+$ can occasionally generate small or unreliable steps, especially when 
	\begin{equation} 
		\frac{ \langle r^{(n-1)}| (\mathcal{\bm{P}}^{-1})^{(n)} | r^{(n)} \rangle }{ \langle r^{(n)}| (\mathcal{\bm{P}}^{-1})^{(n)} | r^{(n)} \rangle } \ll 0, \quad (n \ge 2)
	\end{equation}
	is violated; in that case, the conjugacy condition does not hold. To mitigate this, we adopt the Hestenes-Stiefel (HS) variant with nonnegativity truncation, $\beta = \text{max} (\beta_{\text{HS}}, 0)$, which is known for its strong theoretical properties and practical stability: 
	\begin{equation}
		\beta_{\text{HS}}^{(n)} =  \frac{\text{Re} \langle r^{(n)} - r^{(n-1)}| (\mathcal{\bm{P}}^{-1})^{(n)} r^{(n)} \rangle }{\text{Re} \langle r^{(n)} - r^{(n-1)}| p^{(n)} \rangle }, \quad (n \ge 2).
	\end{equation}
	The HS form measures the component of the new (preconditioned) residual along the update displacement $p^{(n)}$, and thus reacts more directly to curvature information gathered in the latest step. In practice we pair $\text{max}(\cdot, 0)$ with short periodic restarts iterations or when $ \langle r^{(n)}, p^{(n)} \rangle \geq 0 $ to maintain descent and prevent the accumulation of round-off errors.
	
	Because $(\mathcal{\bm{P}}^{-1})^{(n)}$ is symmetric positive definite, both $\beta_{\text{PR}}$ and $\beta_{\text{HS}}$ are well defined and cheap to compute. The use of projected directions $p_i^{(n)}$ preserves normalization by construction and enables a line search along a geodesic on the product-of-spheres manifold (\refsec{section:Optimal step length}). In our benchmarks, HS with restarts delivered the most consistent monotone energy decrease across droplet and supersolid regimes, while PR$^+$ remained competitive in smoother portions of parameter space.
	
	\section{Optimal step length}
	\label{section:Optimal step length}
	
	Ref. \cite{Antoine2017efficient} discusses several implementations of nonlinear CG for \emph{single-component} dipolar BECs. Building on the same philosophy, we design a line search tailored to the \emph{two-component} eGPE and to the manifold constraint $||\psi_{i}||_2 = 1$. Our derivation begins from an unnormalized trial update along the conjugate direction and then restores normalization explicitly.
	
	Let the current iterates be $\psi_{1}^{(n)}$ and $\psi_{2}^{(n)}$, with conjugate direction $d^{(n)}$. We first form the unnormalized step
	\begin{equation}
		{\widetilde{\psi_i}}^{(n+1)}
		= {\widetilde{\psi_i}}^{(n)}
		+\alpha^{(n)} d_i^{(n)}.
	\end{equation}
	Because this update does not preserve $||\psi_{i}||_2 = 1$, we reimpose the constraint by an explicit renormalization,
	\begin{equation}
		\psi_i^{(n+1)} = \frac{\psi_i^{(n)} +  \alpha^{(n)} d_i^{(n)}}{\|\psi_i^{(n)}+  \alpha^{(n)} d_i^{(n)}\|}.
	\end{equation}
	Equivalently, one may view the step as moving along a great circle on the unit sphere in function space; the representation above is convenient for deriving an analytic step size.
	
	To remove the component of the direction parallel to $\psi_i^{(n)}$ and ensure tangency to the normalization manifold, we introduce the rectified direction 
	\begin{equation}
		p_i^{(n)} = d_i^{(n)} - \text{Re} \left\langle d_i^{(n)} \bigg| \psi_i^{(n)} \right\rangle \psi_i^{(n)}.
	\end{equation}
	A small-angle expansion of the normalized state then yields 
	\begin{align}
		&\psi_{i}^{(n+1)} = \left(1- \frac{{\alpha^{(n)}}^2}{2 {\gamma_i^{(n)}}^2}\right) \psi_{i}^{(n)} +  \alpha^{(n)} p_i^{(n)} + \mathcal{O}({\alpha^{(n)}}^3) ,
		\label{eq:New psi normalized}
	\end{align}
	with $\gamma_i = 1 / ||p_i^{(n)}||$. This approximation is used only to obtain a closed-form proposal for the step size; in code, the resulting proposal is evaluated along the \emph{exact} normalized path to guarantee descent.
	
	Substituting \refeq{eq:New psi normalized} into the energy functional \refeq{eq:dimensionless form of the energy functional} and retaining terms through second order produces a quadratic model for the energy variation: 
	\begin{align}
		E^{(n+1)}(\alpha^{(n)}) &\approx E^{(n)} + \alpha^{(n)} \left( E_{1}' + E_{2}' + E_{\text{LHY}}'\right) \nonumber \\
		&+ \frac{{\alpha^{(n)}}^{2}}{2} \left( E''_{11} + E''_{12} + E''_{21} + E''_{22} + E_{\text{LHY}}'' \right).
	\end{align}
	Here the first- and second-order coefficients separate cleanly into mean-field and LHY parts,
	\begin{align}
		E_{i}' &= 2N_i \text{Re} \left\langle p_{i}^{(n)} \bigg| \bm{\hat{H}}_{\text{MF}i}^{(n)} \psi_{i}^{(n)} \right\rangle, \\
		E_{ii}^{''} &= 2N_i \bigg[- \frac{1}{{\gamma_{i}^{(n)}}^2} \left\langle \psi_{i}^{(n)} \big| \bm{\hat{H}}_{\text{MF}i}^{(n)} \psi_{i}^{(n)} \right\rangle  \nonumber\\
		&\quad\hspace{0.9cm} + \left\langle p_{i}^{(n)} \big| \bm{\hat{H}}_{\text{MF}i}^{(n)} p_{i}^{(n)} \right\rangle \nonumber\\
		&\quad\hspace{0.9cm} + \text{Re} \left\langle w_{ii}^{(n)} \big| p_{i}^{(n)} \right\rangle \bigg], \\
		E_{ij}^{''} &= 2N_i \text{Re} \left\langle w_{ij}^{(n)} \big| p_{i}^{(n)} \right\rangle, \quad \left(i \neq j \right),\\
		E_{\text{LHY}}' &= \frac{1}{3 \sqrt{2} \pi^2} \int_0^1 du \sum_{\lambda=\pm} {V_\lambda (u)}^{3/2} {V^1_\lambda (u)},   \\
		E_{\text{LHY}}'' &= \frac{2}{3 \sqrt{2} \pi^2} \int_0^1 du \sum_{\lambda=\pm} \Big[ {V_\lambda (u)}^{3/2} {V^2_\lambda (u)} \nonumber \\
		&\quad \hspace{3.0cm} + \frac{3}{4} {V_\lambda (u)}^{1/2} {V^1_\lambda (u)}^{2} \Big], \nonumber\\
	\end{align}
	with auxiliary quantities 
	\begin{align}
		\gamma_i^{(n)} &= \frac{1}{\|p_i^{(n)}\|} = \frac{1}{\sqrt{\langle p_i^{(n)} |p_i^{(n)} \rangle}},
		\\
		w_{ij}^{(n)}(\bm{r}) &= 2 \Big[G_{ij} \rho_j^{(n)}(\bm{r}) \nonumber \\
		& \hspace{0.7cm} +  D_{ij} \int d\bm{r^{\prime}} \frac{1 - 3 \cos^{2}\theta}{\left| \bm{r} - \bm{r^{\prime}} \right|^3} \rho_j^{(n)}(\bm{r'})\Big]{\psi_j^{(n)}}(\bm{r}), 
		\\
		\rho_k^{(n)}(\bm{r})&=
		\text{Re}\bigg(\psi_k^{(n)}(\bm{r}){p_k^{(n)}}^*(\bm{r})\bigg)\\
		V^1_\pm (u) &= \sum_{i=1}^2 \eta_{ii} n_i' \pm \frac{\delta \delta_1 +  2\eta_{12} \eta_{21} \left( n_1 n_2' + n_1' n_2 \right)}{\sqrt{\delta^2 + 4\eta_{12} \eta_{21} n_1 n_2}}, 
		\\
		V^2_\pm (u) &= \sum_{i=1}^2 \eta_{ii} n_i'' \nonumber \\ 
		&\pm \frac{ \delta_1^2 + 2\delta \delta_2 +  4\eta_{12} \eta_{21} \left(n_1' n_2' + n_1 n_2'' + n_1'' n_2 \right)}{2\sqrt{\delta^2 + 4\eta_{12} \eta_{21} n_1 n_2}} \nonumber \\
		&\mp \frac{\left[ \delta \delta_1 +  2\eta_{12} \eta_{21} \left( n_1 n_2' + n_1' n_2 \right) \right]^2}{2 \left( \delta^2 + 4\eta_{12} \eta_{21} n_1 n_2 \right)^{3/2}} , \\
		n_i' &= 2 \text{Re} \left( \psi_i^{\ast} p_i \right), \\
		n_i'' &= |p_i|^2 - |\psi_i|^2 / \gamma_i^2, \\
		\delta_1 &= \eta_{11}n_1' - \eta_{22}n_2',\\
		\delta_2 &= \eta_{11}n_1'' - \eta_{22}n_2'',
	\end{align}
	and where $ \bm{\hat{H}}_{\text{MF}i}^{(n)} $ denotes the mean-field Hamiltonian without LHY chemical potential $ \mu_{\text{LHY}}^{(i)} $. The terms $w_{ij}^{(n)}$ encode the linearized response of contact and dipolar couplings along the direction $p^{(n)}$, while $V_{\pm}^{1,2} (u)$ capture the first/second variations of the LHY branch potentials.
	
	Minimizing the quadratic model in $\alpha^{(n)}$ gives a closed-form proposal for the step and its componentwise angles, 
	\begin{align}
		\alpha^{(n)} &= - \frac{\sum_{i=1}^{2} E_{i}' + E_{\text{LHY}}'}{\sum_{i=1}^{2}\sum_{j=1}^{2}  E_{ij}''  + E_{\text{LHY}}''} , \\
		\theta_{i}^{(n)} &= \alpha^{(n)} / \gamma_i^{(n)}.
	\end{align}
	In implementation, we evaluate $E(\psi^{(n)} (\theta))$ along the \emph{exact} normalized path and accept the analytic $\alpha^{(n)}$ if it yields a strict decrease of the total energy. If the denominator in the expression above is nonpositive, or if round-off triggers a non-descent step, we revert to a short Armijo backtracking along the same manifold path. This safeguard guarantees monotonic energy descent with negligible overhead and proved effective across droplet, stripe, and supersolid regimes.
	
	\section{Numerical Results}
	\label{section:Numerical Results}
	
	\subsection{Benchmarking Preconditioners and Algorithmic Efficiency}
	\label{section:Benchmarking}
	
	We assess numerical efficiency along two axes: (i) the effect of preconditioning within the conjugate-gradient (CG) scheme, and (ii) a head-to-head comparison between CG and imaginary-time evolution (ITE) under matched discretizations and stopping tolerances. Unless stated otherwise, simulations model a strongly dipolar $^{52}$Cr binary condensate in a harmonic trap with $(\omega_x,\omega_y,\omega_z)=2\pi\times(100,100,800)$ Hz and a magnetic field aligned with $z$. The components are oppositely polarized, $(\mu_1,\mu_2)=(6\mu_B,-6\mu_B)$. All runs use a $256 \times 256 \times 32$ grid and Gaussian initial states perturbed by weak Perlin noise to reduce bias from symmetry and to probe robustness against metastability. In all comparisons we keep the discretization, FFT plans, and stopping criteria identical across methods to ensure fairness.
	
\begin{figure*}[t] 
		\begin{minipage}[t]{.495\textwidth}
			\centering
			\begin{overpic}[width=\linewidth,trim=20 6 6 20,clip]{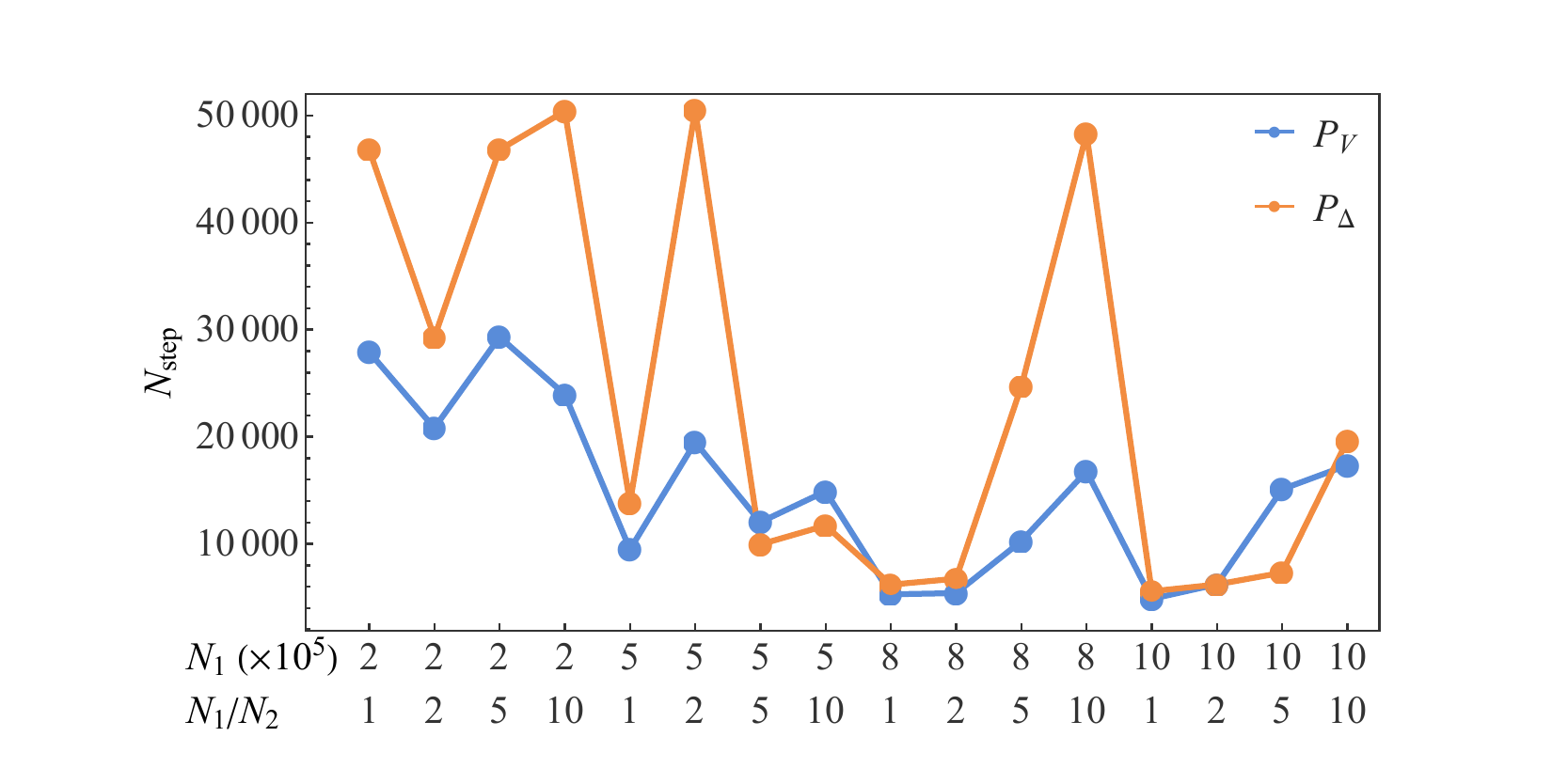}
				\label{subfig:conv_all:a}
				\put(2,47){\small\bfseries (a)} 
			\end{overpic}
		\end{minipage}
		\begin{minipage}[t]{.495\textwidth}
			\centering
			\begin{overpic}[width=\linewidth,trim=20 6 6 20,clip]{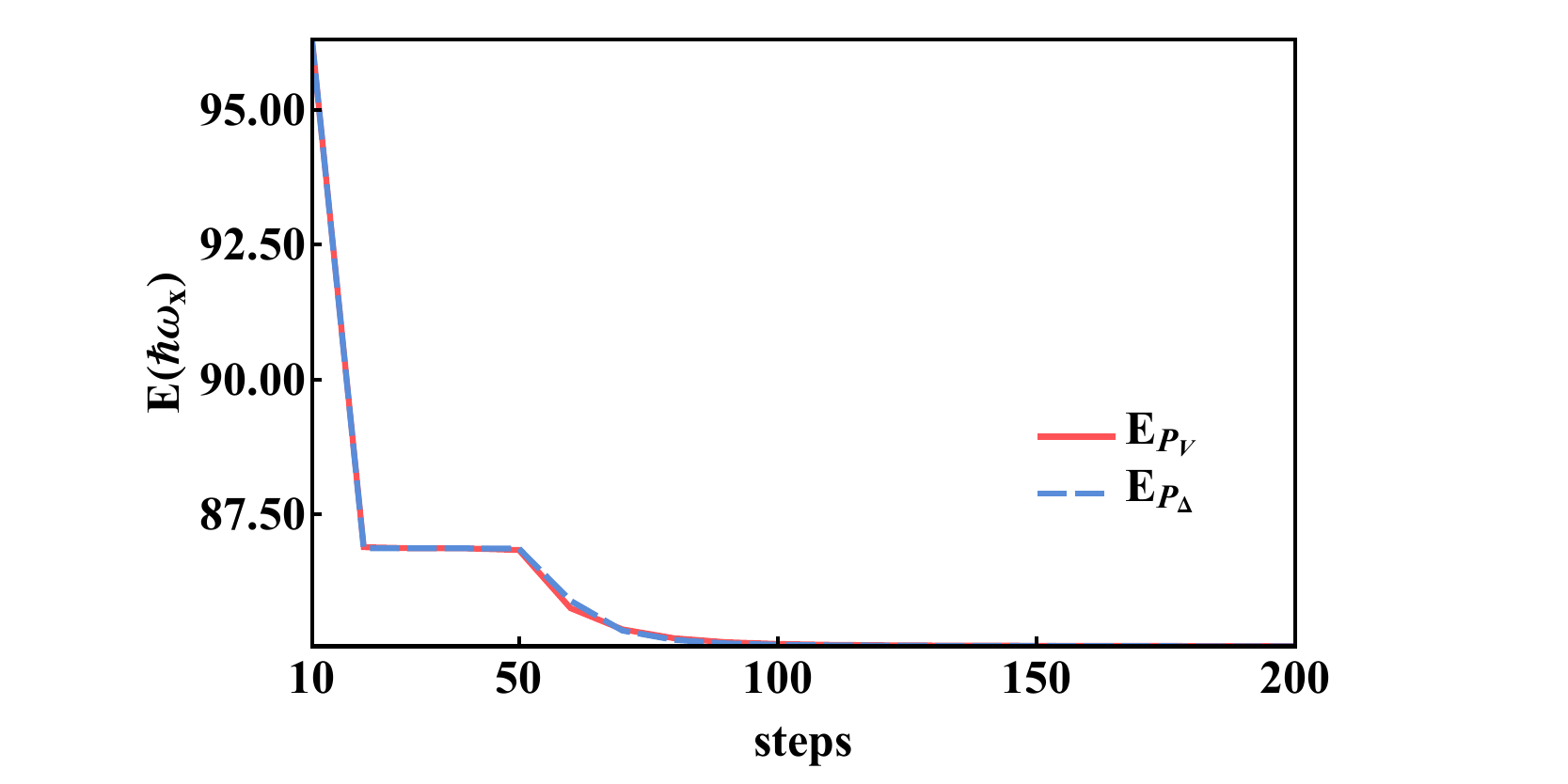}
				\put(2,47){\small\bfseries (b)}
			\end{overpic}
		\end{minipage}
		
		\vspace{2pt} 
		
		\begin{minipage}[t]{.495\textwidth}
			\centering
			\begin{overpic}[width=\linewidth,trim=20 6 6 20,clip]{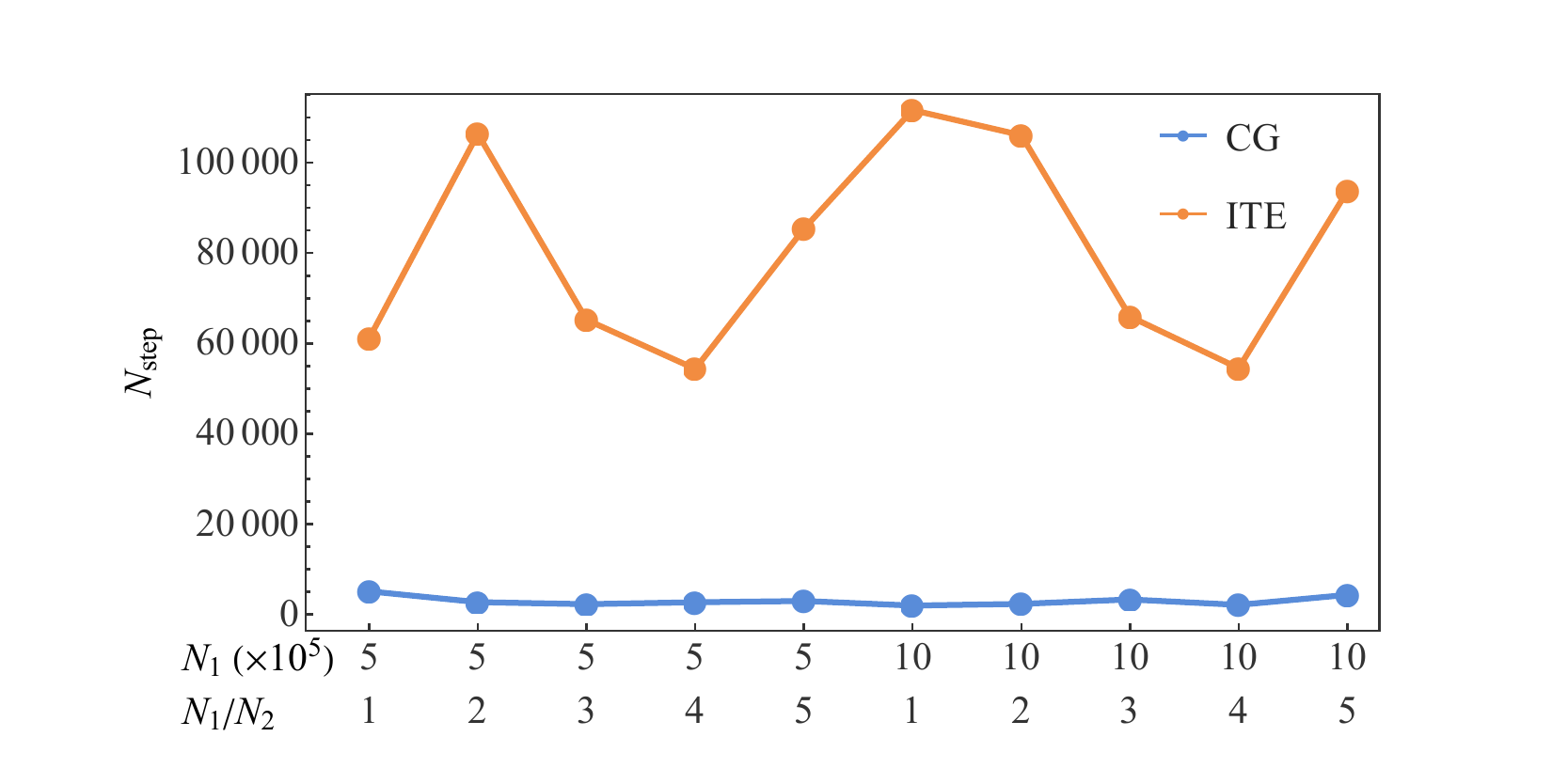}
				\put(2,47){\small\bfseries (c)}
			\end{overpic}
		\end{minipage}
		\begin{minipage}[t]{.495\textwidth}
			\centering
			\begin{overpic}[width=\linewidth,trim=20 6 6 20,clip]{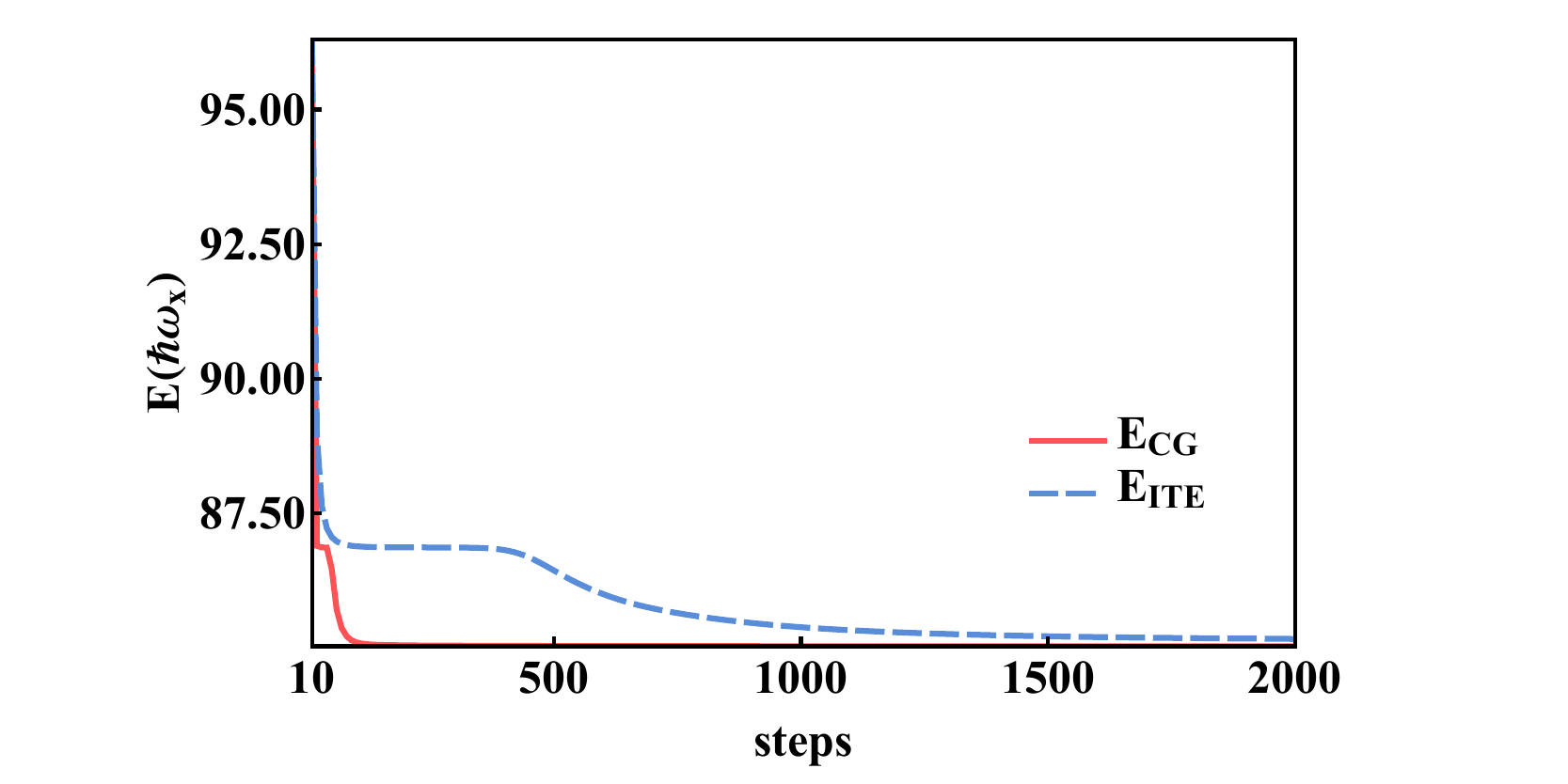}
				\put(2,47){\small\bfseries (d)}
			\end{overpic}
		\end{minipage}
		
		\caption{
			Convergence comparison for a strongly dipolar $^{52}\mathrm{Cr}$ two-component BEC with oppositely polarized components.
			(a) CG iteration counts $N_{\mathrm{step}}$ with preconditioners $\mathcal{P}_V$ and $\mathcal{P}_\Delta$ over $N_1\in{2,5,8,10}\times10^5$ and $N_1/N_2\in{1,2,5,10}$ (x-axis: $N_1$ $(\times10^5)$ on the first line and $N_1/N_2$ on the second; y-axis: $N_{\mathrm{step}}$). $\mathcal{P}_V$ typically converges in fewer steps than $\mathcal{P}_\Delta$.
			(b) Energy decay of CG with $\mathcal{P}_\Delta$ and $\mathcal{P}_V$ at $N_1=10\times10^5$ and $N_1/N_2=1$; convergence is declared when the maximum residual falls below $10^{-10}$.
			(c) Iteration counts of CG and ITE for $N_1\in{5,10}\times10^5$ and $N_1/N_2\in{1,2,3,4,5}$ (same axis labeling as in (a)); for fairness, convergence for both methods is defined by an energy variation below $10^{-6}$.
			(d) Energy decay at $N_1=10\times10^5$ and $N_1/N_2=1$, highlighting the markedly faster convergence of CG compared to ITE.
		}
		\label{fig:conv_all}
	\end{figure*}

	For the preconditioner study, CG convergence is declared when the maximum residual falls below $10^{-10}$. \reffig{fig:conv_all}(a) reports iteration counts for $\mathcal{P}_\Delta$ and $\mathcal{P}_V$ over $N_1\in{2,5,8,10}\times10^5$ and $N_1/N_2\in{1,2,5,10}$, and \reffig{fig:conv_all}(b) shows representative energy decay at $N_1=10\times10^5$ and $N_1/N_2=1$. Across the entire sweep, $\mathcal{P}_V$ systematically reduces the number of iterations relative to $\mathcal{P}_\Delta$, consistent with the expectation that a real-space (Hessian-inspired) scaling is more effective when local interaction and trap terms dominate the stiffness.
	
	To compare with ITE, we adopt an energy-based stopping rule—termination once the absolute energy variation between successive iterations falls below $10^{-6}$—for both solvers. This criterion avoids penalizing ITE with an excessively strict residual target while still ensuring comparable accuracy. As shown in \reffig{fig:conv_all}(c,d), CG converges substantially faster than ITE throughout the tested parameter range; the gain is often one to two orders of magnitude in iteration count. In all cases, the same initial conditions and perturbations are used for the two solvers.
	
	\reftab{Table:Ground State Energies} summarizes the ground-state energies per particle obtained by the two approaches. Agreement is uniformly close, with CG returning slightly lower energies in most cases. This is consistent with the observation that, in landscapes populated by many nearly degenerate local minima, CG more readily reaches lower-energy branches at matched tolerances.
	
	\begin{table}[t]
		\begin{ruledtabular}
			\centering
			\caption{Ground State Energies \label{Table:Ground State Energies}}
			\begin{tabular}{l c c c} 
				$N_1 \times 10^5$      
				& $N_1/N_2$
				& $E_{\text{CG}}/N (\hbar \omega_x)$
				& $E_{\text{ITE}}/N (\hbar \omega_x)$\\
				\midrule
				5   & 1 & 65.911447 & 65.945427 \\
				5   & 2 & 59.381773 & 59.388998 \\
				5   & 3 & 56.615889 & 57.256970 \\
				5   & 4 & 55.645250 & 56.260503 \\
				5   & 5 & 55.106218 & 55.704514 \\
				10  & 1 & 86.440507 & 86.481219 \\
				10  & 2 & 77.888680 & 77.897386 \\
				10  & 3 & 75.157023 & 75.160370 \\
				10  & 4 & 73.907971 & 73.916195 \\
				10  & 5 & 73.219556 & 73.229081 \\
			\end{tabular}
		\end{ruledtabular}
	\end{table}
	
	In all experiments, we fixed the discretization and solver tolerances across methods and varied only the preconditioner or the algorithmic class (CG vs ITE). Random seeds for initial noise were held constant when comparing methods for a given parameter set. These choices remove confounding factors and attribute performance differences to the algorithms themselves.
	
	\subsection{Computational cost}
	\label{subsection:Computational cost}
	
	The dominant expense per iteration stems from FFT-based operations (kinetic updates and dipolar convolutions); pointwise nonlinearities are $O(N)$ and comparatively negligible. For the ITE scheme, one iteration requires 8 FFTs (forward/backward). In contrast, CG triggers additional transforms to evaluate the quadratic line-search model and to maintain conjugacy, yielding 34 FFTs per step without preconditioning or with $\mathcal{P}_V$, and 38 FFTs with $\mathcal{P}_\Delta$. The per-step costs can therefore be summarized as
	\begin{equation}
		C_{\mathrm{ITE}}^{\text{step}} \sim 8\, O(N \log N), \qquad
		C_{\mathrm{CG}}^{\text{step}} \sim m\, O(N \log N),
	\end{equation}
	with $m=34$ or $38$ depending on the preconditioner. Although the CG step is thus more expensive by a factor of $\sim4$–$5$, \reffig{fig:conv_all}(c) shows that CG requires far fewer steps to reach the same accuracy, so the \emph{total} wall-clock time is substantially reduced. Among the tested options, $\mathcal{P}_V$ consistently provides the best time-to-solution, reflecting its stronger conditioning of the low-$k$ stiffness induced by trap and interactions.

	\subsection{Consistency check with previous studies}
	
	We next verify the solver against representative results from Refs.~\cite{Santos2023self,Zhang2024metastable}. Unless otherwise stated, runs use a $128 \times 128 \times 64$ grid with periodic boundaries in the $x$-$y$ plane and Gaussian initial conditions perturbed by weak noise. CG is used throughout, with convergence declared at a maximum residual below $10^{-10}$; dipole orientations and interaction parameters follow the cited works.
	
\begin{figure*}[t]  
		\centering 
		\includegraphics[width=0.75\textwidth]{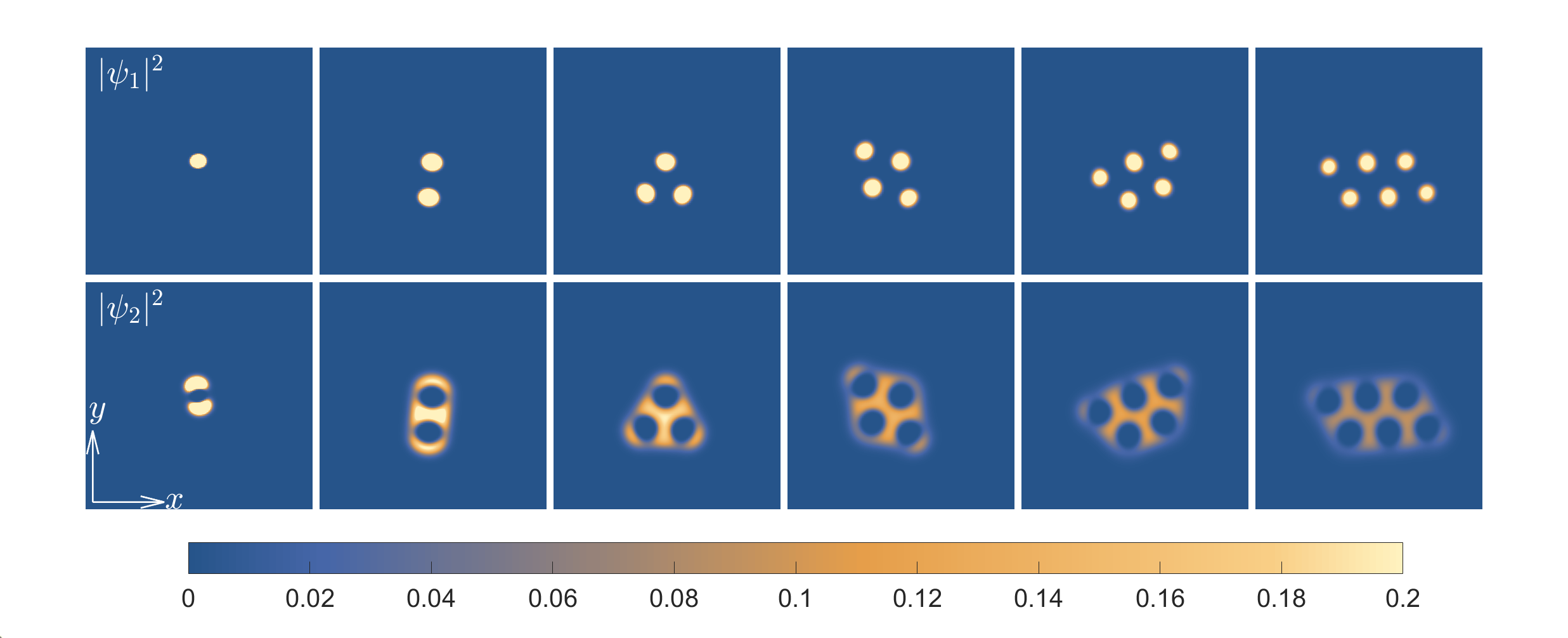}  
		\caption{
			Numerical results obtained using the CG method, showing self-organized patterns of a polarized dipolar BEC with $N_1=N_2=3\times10^4$ $^{162}$Dy atoms in a $\hat{\bm z}$-aligned magnetic field.
			The trapping frequencies $\omega_z/2\pi={100,250,300,350,370,425.3}$,Hz generate droplets with $N_d={1,2,3,4,5,6}$, respectively.
			Parameters: $\mu_1=-\mu_2=10\mu_\text{B}$, $a_{11}=50a_0$, $a_{22}=70a_0$, $a_{12}=150a_0$.
			Simulations use a $128\times128\times64$ grid with periodic $xy$ boundaries. Results reproduce key features of Ref.~\cite{Santos2023self}.} 
		\label{fig:repre_patterns} 
	\end{figure*}

	For a counter-polarized $^{162}$Dy mixture in a $\hat{z}$-aligned field with $N_1 = N_2 = 3 \times 10^{4}$, we reproduce the droplet-number sequence reported in Fig.~3 of Ref.~\cite{Santos2023self}: choosing $\omega_z / 2 \pi = 100, 250, 300, 350, 370, 425.3$ Hz yields $N_d = {1, 2, 3, 4, 5, 6}$, respectively. The corresponding density patterns obtained with CG are shown in \reffig{fig:repre_patterns}. These results confirm that the present implementation captures the transition from single to multiple self-bound droplets as the axial confinement is strengthened.
	
\begin{figure*}[t]  
		\centering  
		\includegraphics[width=0.8\textwidth]{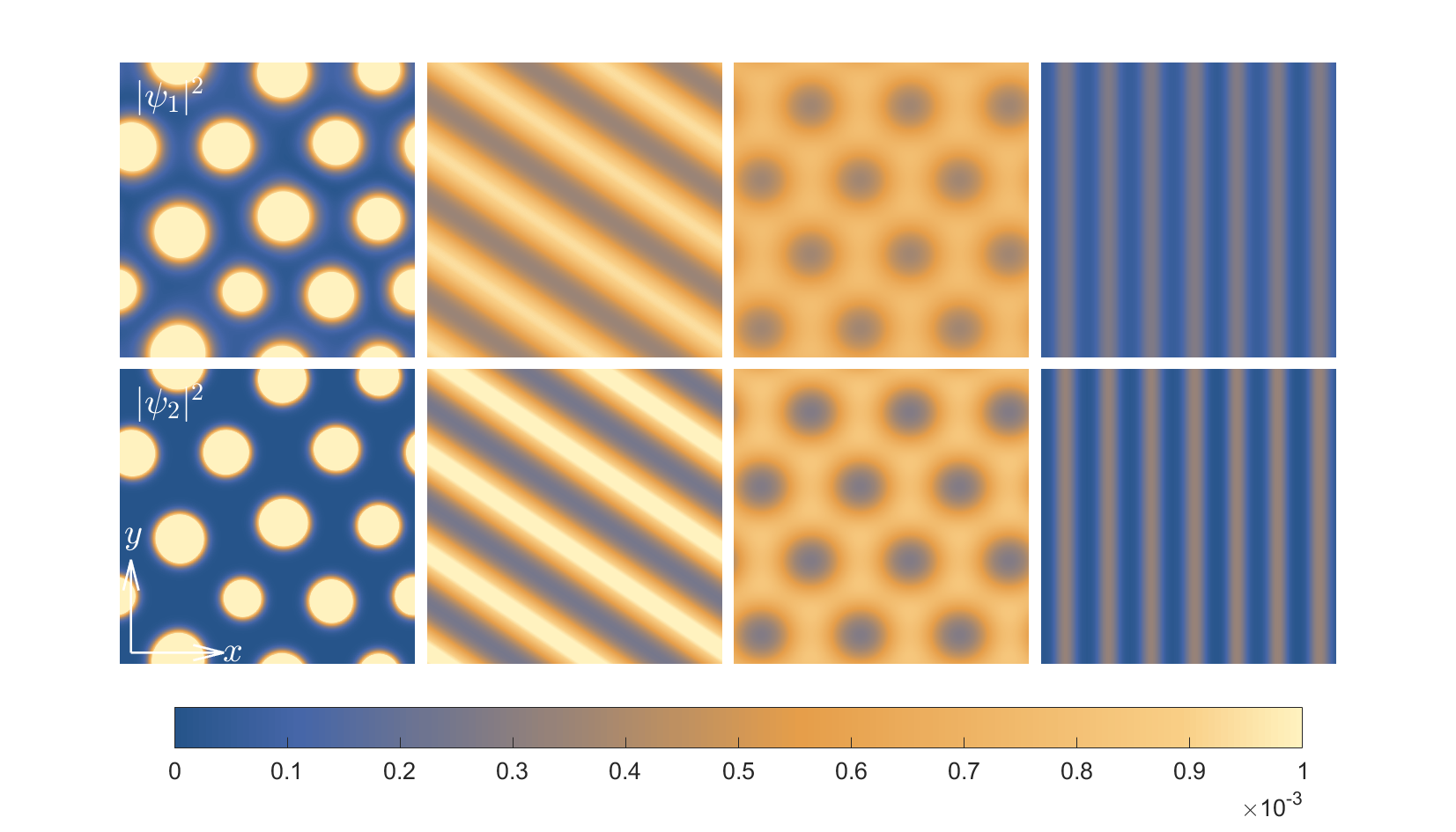} 
		\caption{
			Numerical reproduction of Ref.~\cite{Zhang2024metastable} for Dy–Er mixtures.
			Parameters: $\omega = 2 \pi \times 900 $ Hz ($\omega_z = 0.08$ dimensionless); dipole moments $\mu_1 = 10 \mu_B$, $\mu_2 = 7.07 \mu_B$.
			Balanced case with $N_1 = N_2 = 10^6$ and $\rho_1 = \rho_2 = 625$, using $a_{11} = 118.5 a_0$, $a_{22} = 58.95 a_0$, and $a_{12} = {46.49,57.65,58.5} a_0$.
			In another configuration ($N_1 = N_2 = 4 \times 10^6$, $a_{11} = 171.6a_0$, $a_{22} = 85.15 a_0$, $a_{12} = 16.7375 a_0$), the parameters expected to produce a ring state instead lead to a stripe structure.
		} 
		\label{fig:Zhang}  
	\end{figure*}

	We further consider Dy-Er mixtures as in Ref.~\cite{Zhang2024metastable}. For panels (a) - (c) of \reffig{fig:Zhang}, a balanced system with $N_1 = N_2 = 10^6$ (nondimensional density $\rho_1 = \rho_2 = 625$) is simulated using $a_{11} = 118.5,a_0$, $a_{22} = 58.95,a_0$, and $a_{12} = a_{21} \in {46.49,57.65,58.5},a_0$. For panel (d), parameters $N_1 = N_2 = 4 \times 10^6$, $a_{11} = 172.9,a_0$, $a_{22} = 85.8,a_0$, and $a_{12} = a_{21} = 15.0,a_0$ (cf. Ref.~\cite{Zhang2024metastable}) produce a stripe rather than a ring texture in our computation. This deviation is consistent with an energetic preference for stripes at the given settings and illustrates the sensitivity of multistable patterns to initial conditions, box size, and boundary conditions; CG tends to select the lower-energy branch at matched tolerances. Overall, the reproduced patterns and transitions align with the qualitative and quantitative trends reported previously.
	
	\section{Conclusion}
	\label{section:Conclusion}
	
	We have introduced a mixture-aware, preconditioned nonlinear conjugate-gradient solver for ground states of binary dipolar Bose-Einstein condensates governed by the extended Gross-Pitaevskii energy with LHY corrections. In contrast to single-component CG schemes, our method (i) formulates the minimization on the product-of-spheres manifold and updates in its tangent space, thereby enforcing two normalization constraints with interspecies couplings; (ii) performs a manifold-preserving analytic line search derived from a quadratic energy model that explicitly includes cross-component mean-field terms, nonlocal dipolar interactions, and LHY variations, and validates the step along the exact normalized path; and (iii) deploys complementary preconditioners, specifically a real-space diagonal (Hessian-inspired) scaling and a Fourier-space kinetic scaling, applied in sequence to balance low- and high-$k$ stiffness. Robust direction updates (Hestenes-Stiefel $\beta$ with restarts and a $\beta \to 0$ safeguard) ensure monotone energy descent in multistable landscapes (droplet/stripe/supersolid).
	
	Head-to-head benchmarks on matched grids and tolerances show that, although a CG step incurs a larger transform budget (34–38 vs 8 FFTs for ITE), the solver reduces iteration counts by one to two orders of magnitude and thus wins in time-to-solution, while typically reaching slightly lower energies, indicating improved resilience to metastability. Among tested options, the diagonal preconditioner $\mathcal{P}_{V}$ provides the strongest conditioning and the best overall performance.
	
	From the physics side, the framework reproduces representative textures and droplet-stability windows reported for dipolar mixtures, consistent with the expected contraction of the miscible window. These results support the method’s reliability for large-scale parameter scans and phase-boundary mapping.
	
	Practically, the solver is modular and reproducible: preconditioners act block-diagonally by component; nonlocal operations are amortized via shared FFTs; and the line search reuses quantities already computed for the residual and energy model. Together with strict normalization and monotonicity safeguards, these features make the approach well suited to large three-dimensional grids and regimes with many low-lying minima.
	
	Natural extensions include adaptive/blended preconditioning (dynamic weighting of $\mathcal{P}_{V}$ and $\mathcal{P}_{\Delta}$), numerical continuation for efficient phase-diagram sweeps and bifurcation tracking, Bogoliubov–de Gennes analysis for excitations and stability, and GPU/NUFFT acceleration or adaptive meshes for extreme aspect ratios.
	
	In summary, the mixture-aware innovations, namely a manifold-preserving analytic line search with dipolar and LHY cross terms, sequentially composed preconditioning tailored to coupled stiffness, and robust CG updates, establish a reliable, efficient, and scalable tool for quantitative studies of multicomponent dipolar quantum superfluids and for quantitatively connecting numerical metastable branches to experimentally accessible states.
	
	\begin{acknowledgments}
		We acknowledge helpful discussions with Jens Hertkorn and Mingyang Guo.
		K.-T.X. was supported by the MOST of China (Grant No. G2022181023L) and NUAA (Grant No. YAT22005, No. 2023YJXGG-C32 and No. XCA2405004). W.B. was supported by NUAA (Grant No. 202510287091Z).
	\end{acknowledgments}
	
	\bibliography{references(final)}
	
\end{document}